Łukasz W. Niparko Ph.D.

# Bowling Online:

## Accounting for Civil Society Reshaped
## into Streamlined Photons within a Fiber Network

**Abstract**

Civil society has been deemed by various scholars, such as Robert D. Putnam, to be a predictor and a cornerstone of a robust and consolidated democracy (Putnam et al., 1993). Putnam highlights in his book *Bowling Alone* (2000) that American civil society has become weaker: people organize less, and literally, they *bowl alone*. But what if there is yet another aspect to Putnam's story that has not been fully accounted for, namely the rise of Digital Civil Society (DCS)? Perhaps people in the third decade of the 21st century *bowl online*. They still organize, mobilize, and care for their civil liberties and democratic institutions; however, the public sphere in which this takes place has shifted online to cyberspace (Bernholz et al., 2013) or to what still needs to be conceptualized—the digital public sphere (DPS), which this article attempts to measure and demarcate.

**Keywords:** digital civil society, digital public sphere, sector Pi, survey research

***"Do not be inactive!"***—Alexei Navalny[1]

## Introduction

In this article, I attempt to conceptualize, operationalize, and measure digital civil society (DCS) through a mixed methodology that ensures the robustness of this study and helps to account for the complexity of DCS and its extent. I answer the following question(s): Is DCS becoming the *new default* form of civil society, and what are the scope and volume of DCS participation among surveyed individuals in the United States (US)?

The structure of this article is as follows: first, I define DCS and digital public sphere (DPS) and discuss the etiology of DCS and its relation to *traditional* civil society. Then, I place the concept of DCS in the context of the third sector, the systems theory, the idea of Jürgen Habermas's public sphere (1991), and the idea of the Sector π (Pi) by Jan Sowa (2015). Next, I introduce the mixed methods used in this project: the collection and analysis of representative survey data, along with semi-structured open-ended interviews. Finally, I present my findings and discuss their implications.

## *Definitions*

DCS is one of the cornerstones of digital democracy and netizenship (that is, empowered agency within the digital sphere). Netizens can be defined as "those who use the Internet regularly and effectively [, and] who use technology for political information to fulfill their civic duty" (Mossberger et al., 2007). The concept of DCS is still uncharted territory for many in social sciences. Lucy Bernholz, a pioneering scholar of DCS defines it as: "the ways, we use private resources to organize, create, distribute, and fund public benefits in the digital age" (Bernholz, 2013). I define DCS as the occurrences and mobilizations of civil society that take place online, e.g., on social media, web forums, and through other online communication technologies, or that initiate online and extend into the offline (public) sphere. Simply, it is a civil society with the epithet *digital*. The very civil society that also comes with new challenges and opportunities such as data handling and data donations, "a non-commercial transfer of data" (Prainsack, 2019). Finally, Larry Diamond, defines *traditional* civil society as "the realm of organized social life that is voluntary, self-generating, (largely) self-supporting, autonomous from the state, and bound by a legal order or set of shared rules" (Diamond, 1994, p.4).

## *What is DCS?*

Civil society has been deemed by various scholars, such as Robert D. Putnam, to be a predictor and cornerstone of a robust and consolidated democracy (Putnam et al., 1993). Putnam highlights in his book *Bowling Alone* (2000) that American civil society has become weaker: people organize less, and literally, they *bowl alone*. Thus, democracy can quickly erode. Indeed, the global democratic erosion (also called "backsliding" or the "third wave of autocratization") has become a 21st-century occurrence, spanning from Brazil to the Philippines and within the US (Bermeo, 2016; Lührmann & Lindberg, 2019). In fact, today we may be living in a period of chronic democratic erosion. But what if there is yet another aspect to Putnam's story that has not been fully accounted for, namely DCS? Perhaps people in the third decade of the 21st century *bowl online*. They still organize, mobilize, and care for their civil liberties and democratic institutions. However, the public sphere in which this occurs has shifted online to the cyberspace (Bernholz et al., 2013) or to what still needs to be conceptualized—the DPS. I define DPS as a virtual space of unrestricted interpersonal interaction and/or as an accessible space *per se* that may, but does not necessarily have to, be used by netizens for diverse purposes in accordance with the law, or when necessary, as a site of civil disobedience.

---

[1] Alexei Navalny (1976-2024), a Russian dissident, was a prominent advocate for DCS, particularly given the limitations of traditional civil society (TCS) within the Russian Federation. The Internet provided him with a platform to reach millions of viewers and gain thousands of followers.

After all, civil society has helped bring down totalitarian regimes, accelerated democratization in many instances, and contributed to the rise of new repressive governments, as with the account of the Weimar Republic and the Third Reich by Sheri Berman (1997). Civil society effectively achieves societal goals, especially nonviolent ones (Chenoweth & Stephan, 2011). It also denotes a form of cultural capital among individuals and nations and is associated with the prospects of stable democracy (Tusalem, 2007). In the 2020s, civil society is also digital. In many parts of the world, the *scape* of Jürgen Habermas (offline) public sphere has shifted into Jaron Lanier's virtual reality (Habermas, 1991; Lanier, 2014), or rather to a digital public sphere/cyberspace, as 47.1% of my survey respondents indicated that they actively participate in the DCS (Table 1).

**Table 1**

Are you part of any online group, including those on social media (please specify)? (Data from a US-based representative sample conducted on February 25th, 2024).

| response | count | % |
|----------|-------|------|
| yes | 164 | 47.1 |
| no | 184 | 52.9 |

For many, the digital journey of civil society began on January 1st, 1994, with the anti-North American Free Trade Agreement movement of the Zapatistas in Chiapas, Southern Mexico, which published its manifesto online (Kowal, 2002). From there, DCS grew and continued its journey with more or less successful campaigns that stirred local and global attention, as some viewed DCS as a new source of hope while others dismissed it as mere *slacktivism* (Reardon, 2013). *Slacktivism* refers to minimal effort and non-committal support for social causes, often through social media actions such as likes, shares, and online petitions (Rotman et al., 2011). After all, DCS's influence can be ephemeral or long-lasting (just as is the case with *traditional* civil society), as with the legacy of the Human Rights Campaign, which was instrumental in raising awareness of marriage equality in the US for same-sex couples (Vie, 2014). DCS played an indispensable role in mass protests, e.g., against ACTA (Mercea, 2017), during Occupy Wall Street (Gleason, 2013), and within the Black Lives Matter (Mundt et al., 2018) movements; it generated various petitions and campaigns through *ONE Campaign* or *Avaaz* (Aragón et al., 2018), and spurred online fundraising efforts in support of diverse causes (Zhang et al., 2021). What is more, the story of civil society activism shifting online and/or into grassroots movements may also reflect inevitable non-governmental organization (NGO) *fatigue*, as some become disillusioned or disinterested in institutionalized, hierarchical, and donor-dependent civil society organizations (CSOs) (Alvarez, 2009; Roche & Hewett, 2013; Roy, 2014). It also produced projects that failed to withstand the test of time and proved not effective as intended (beyond garnering attention), like the "Kony 2012" campaign to capture the Ugandan warlord, Joseph Kony (Bal et al., 2013). Many DCS initiatives have originated or accelerated digitally and then moved to the analog public sphere, where protests, sit-ins, and other forms of dissent take place. Nonetheless, the critical question remains whether there is a requirement for offline *follow-up* for the efficacy of DCS; or whether digital means alone are, or will be, sufficient to achieve civil society's intended goals, such as influencing a particular policy, achieving leverage over the people in power, monitoring them, delivering services, crowdsourcing people and/or resources, or promoting awareness and engaging in advocacy.

### *Contextus*—it means to weave together

Below, I will provide the theoretical context for DCS. As I learned from my mentors, in Latin, *contextus* derives from *contexere*, meaning to weave together. Specifically, I explore the resilient nature of civil society beyond its traditional classification as the so-called "third sector," its relationship to systems theory, and its grounding in the pursuit of a robust democracy.

### *Sector π*

Traditionally, civil society has been regarded as the third sector, distinct from both private and governmental entities. Sociologist Jan Sowa offers a thought-provoking perspective by describing the third sector as the "$π$ sector" (2015). Sowa's $π$ sector represents the diversity and robustness of the non-governmental and non-private world. The concept of Sector $π$ applies well to DCS, as it encompasses various forms, structures, and meanings. Sowa highlights the significance of Putnam's work, which, when considered alongside Pierre Bourdieu's theories, underscores the role of social capital in shaping societal dynamics. Bourdieu (1977) emphasizes how social capital contributes to the accumulation of wealth across generations, while Putnam et al. (1993) illustrate its impact on reinforcing regional divides, such as those between Northern and Southern Italy. Bourdieu perceives social capital as, "the aggregate of the actual or potential resources which are linked to possession of a durable network of more or less institutionalized relationships of mutual acquaintance and recognition" (Bourdieu, 1985). Putnam builds on this notion by inserting social and cultural capital into his broader framework of civil society and democracy. After all, the story of the public sphere is also one of the creation and accumulation of social and cultural capital; first through education, and then through humanistic exchange, where individuals encounter one another within the public sphere, as outlined by Habermas (1991).

Habermas traced the structural transformation of the public sphere to the wave of Enlightenment (1991). The public sphere, as conceptualized by Habermas, emerged as a space where individuals could engage in critical debate about matters of public concern. This development occurred alongside various social institutions, including educational ones, and was shaped by the rise of coffeehouses, salons, and print media in the 18th-century. Today, one does not need to be a socialite to access the leading ideas dispensed on one of the largest DCS projects, *Wikipedia*, or those widely available through platforms such as *TED Talks*, *Khan Academy*, or *Coursera*. Or can access new forms of knowledge and creation enabled by generative artificial intelligence (AI). Today, the cost of one coffee can grant hours of engagement within the digital sphere. DCS comes with a barrier-breaking force where social movements can go beyond urban space into a global infosphere: the environment where information is produced, processed, stored, and exchanged (Floridi, 2009). Thus, civil society participation can be more accessible, especially in small towns and rural areas. A phenomenon that is already observable in data, as shown in Table 2, which reports levels of DCS participation by place of residence in the US. Rural communities (with a population of less than 10,000) ranked third (out of five) among the places of domicile of the DCS participants. In this sense, we observe an extension of opportunity to participate for diverse range of people in civil society beyond those who have access to the urban public space or nonprofits. Nonetheless, we need to keep in mind the access limitations faced by individuals excluded from digitalization due to socio-economic barriers and other factors (Van Dijk, 2020). Nonetheless, Habermas (1991) points out that modernization of which digitalization is a continuation can introduce significant obstacles to the public sphere, particularly by transforming citizens into passive consumers. In his later work, Habermas (2006) remains skeptical about digitalization's potential to strengthen the public sphere, expressing concerns about its role in fostering fragmentation and weakening deliberative democracy. In the Global North, the digital sphere comes with many limitations inherent to neoliberal capitalism, since corporations control many aspects of it (Lanier, 2010; Zuboff, 2017); it also comes with the exclusion of certain users and restrictions on their speech.

**Table 2**

Summary table comparing TCS (Traditional Civil Society), DCS, and PDCS (Political Digital Civil Society) scores and place of inhabitance. (Data from a US-based representative sample conducted on February 25th, 2024).

| | average of TCS_score | average of DCS_score | average of PDCS_score |
|---|---|---|---|
| rural area (less than 10,000 people) | 6.70 | 2.96 | 2.00 |
| suburban area (surrounding an urban area) | 6.13 | 2.75 | 1.78 |
| town or small city (population between 10,000 to 500,000 people) | 7.31 | 2.95 | 2.01 |
| urban area (city with a population of more than 500,000 people) | 6.83 | 3.83 | 2.29 |
| large city (city with a population of more than 1 million people) | 8.40 | 3.79 | 2.36 |
| *average* | *6.96* | *3.10* | *2.02* |

Therefore, the problem of digital civil society is not only one of the DPS and its ownership, but also one of how digitalization can spur and maintain social capital. This shift carries enormous potential for democratizing social and cultural capital through greater accessibility to the digital sphere. The growth of a person through the accumulation of social and cultural capital online leads to the development of the netizen, a person not only able to consume digital content but also able to produce it and influence reality beyond it. Ultimately, what is at stake for DCS is the formation of a community of netizens who possess genuine agency in achieving societal goals.

***DPS and Systems Theory***

In North America, DCS remains reliant on big tech to provide the infrastructure necessary for its functioning. This dependence stems from the fact that society has yet to fully delineate the DPS, determine its ownership, or establish a concrete framework for its governance. Moreover, significant debate persists regarding the scope and boundaries of the DPS (Schäfer, 2015). We stand before a choice: 1). accept the current *status quo*, in which that space is owned by private tech companies and/or governments; or 2). make a clear demand for what the DPS should look like, modeled on analog public spaces such as squares, parks, public libraries, or cooperatives. Currently, the fates of DCS and the DPS are left to market forces and potential (but unlikely) regulation. Thus, very salient became the words of one of my interlocutors, who highlighted that "we are the regulators" (Participant #8, personal communication, March 24, 2024).

While discussing the theoretical assumptions of DCS and DPS, we cannot omit the Socio-Technical Systems (STS) Theory, which focuses on the interdependence between social and technical systems and people's presence as agents within them (Sony & Naik, 2020). STS theory, "assumes that the system can be understood from its interactions between the various parts of the system" (Sony & Naik, 2020, p.4). Social media serve as a prime example of STS theory applied in reality (Lombardo et al., 2021). Fred E. Emery and Eric L. Trist, who were instrumental in developing STS theory, emphasized its decentralized nature (Emery & Trist, 1960). Decentralization in the digital sphere brings with it a degree of democratization. DCS represents yet another materialization of STS, where discourse and engagement occur within digital infrastructures. Still, DCS operates with certain clear limitations on how we can exercise our agency. After all, we cannot (*yet*?) chain ourselves, set up an encampment, or squat in a place that is not physical.

In addition to STS, social network analysis can be used to understand DCS. For example, Molly McLure Wasko and Samer Faraj combined social network analysis with cultural capital theory (2005). Since DCS is both a discursive and action-oriented landscape, the insights drawn by Lombardo et al. (2021) from Wasko and Faraj's study are particularly helpful in explaining why people come together within DCS:

> [1.] **individual benefit**, expected by a member from his/her contribution, even without direct acquaintance and without mechanisms encouraging reciprocity; online and professional reputation is often one of the most important expected benefits; [2.] **structural capital**, number and strength of connections and the associated habit of cooperation and propensity for collective actions; [3.] **relational capital**, personal relationships and trust among members, perceived commitment and reciprocity, acceptance of common norms; [4.] **cognitive capital**, required for any meaningful interactions between members, in the form of basic shared understanding and semantics, including contexts, norms, languages, interpretations and narratives (pp. 4-5).

Just as one atom of oxygen and two atoms of hydrogen merge to form a new substance, people who come together digitally transform from networks of independent customers into organized nodes within a system, one that can grant them netizenship, provided they possess the awareness and inclination to pursue it.

***DCS as democracy's litmus paper***

There is a salient nexus between the digitalization of civil society and democracy. In less democratic regimes, governments attempt to limit the public sphere for civil society (e.g., Alscher et al., 2017). As laws curtailing civil society come into effect, DCS emerges as a new *agora* for mobilization and a potential means of upholding democratic values. For example, Poland, between 2015 and 2023, under the rule of the Law and Justice party (*Prawo i Sprawiedliwość*), became an arena of pro-democratic DCS mobilization, where citizens could assume the role of netizens and unite in defense of democratic values through initiatives such as Action Democracy (*Akcja Demokracja* in Polish). *Akcja Demokracja* leveraged digital tools and grassroots campaigns to spread awareness of the dangers of democratic erosion and to pressure those in power by demonstrating mass mobilization among engaged netizens, voicing concerns over the decline of the rule of law and the erosion of certain rights (*Akcja Demokracja*, n.d.).[2]

A social scientist and feminist, Carole Pateman, highlighted the importance of gender in ensuring the success of the social contract and democracy, in *Participation and Democratic Theory* (1970), where she highlighted the necessity of *"we the people"* participating in the democratic process at every level of *polity*. The strength of institutions depends just as much on skilled bureaucrats as on informed and active people who enable institutions to function by grounding them in checks and balances, social ethics and public oversight. Likewise, Benjamin Barber envisions strong democracy as participatory politics (2003). Such democracy can be strong not only through the power of representative democracy *per se*, but through a living democratic process that might begin on your sofa with a laptop and culminate in holding those in power accountable, or even becoming one of them through self-empowerment, whether as part of a community or as a frontrunner. More importantly, it is crystalized in the role of the individual as a digital agent of democratic *praxis*: the netizen.

Additionally, there is another tangent to the scope of DCS—digital totalitarianism/unfreedom (Diamond, 2019; Risse, 2021, 2023). It is exemplified by the regimes using digitalization as a tool to restrain already scarce liberties or groups such as the *QAnon* that also mobilize online yet for an anti-democratic agenda (Cover et al., 2022). DCS is not free from attempts to limit its operation and curtail mobilization. For instance, countries like Iran suppress DCS when pro-democratic online movements stand in opposition to (digital) totalitarianism (Diamond, 2019). Likewise, eroding democracies may impose digital surveillance tools, such as *Pegasus*, to intimidate civil society (Citizen Lab, 2018). Various non-democratic or less-democratic places utilize digital means to operate their regimes, e.g., as illustrated by Viola Rothschild's (2024) work tracing the procurement of digital surveillance technologies.

Thus, conceptualization, operationalization, and measurement of DCS could become a roadmap for democratic societies in not only embracing DCS but also being able to remain democratic (through netizenship) when confronted with digital-totalitarian competitors, or internal forces that erode democracy and human rights, or barriers that inhibit the empowerment of Internet users.

---

[2] There have been public concerns raised about *Akcja Demokracja*'s use of social media content, with some critics suggesting that foreign sponsorship may have influenced its normatively questionable role in the Polish presidential election of 2025 (e.g., Jadczak & Słowik, 2025).

## Methods

In this article, I answer the following question(s): Is the DCS a new default for civil society? And what is the scope and volume of DCS participation among surveyed persons? It is an explanatory sequential study[3] (Creswell & Plano, 2018; Ivankova et al., 2006). The quantitative phase encompasses original survey data, while its qualitative part consists of in-depth semi-structured open-ended interviews. The interlocutors were recruited using the so-called "snowball-sampling" technique (Forman et al., 2008), also called the respondent-driven-sampling technique, where each and every participant had a chance to influence the qualitative data collection (Boswell & Babchuk, 2023). Through this research, I engaged 364 participants. In Table 3, I describe the research instruments I used. However, I would also like to acknowledge those who, for different reasons, including socio-economic barriers, are affected by digital exclusion and could not participate in this data collection as they operate on the margins of digitalization. Their perspectives should not be overlooked.

**Table 3**

Research Instruments.

| instrument | sample size | number of questions | specification |
|---|---|---|---|
| Digital Society: opinions, attitudes, and participation *(Part 1: **Digital Civil Society**)* | 348 | 37 | demographics; opinions, attitudes, and participation in digital and offline civil society; political views |
| Interview Questions for Academic, Governmental, Industry and Civil Society Stakeholders | 16 | 24 | personal definitions and viewpoints on concepts such as civil society, democracy, and digitalization; discussions of the societal and individual implications of rapid technological advancements; and prompts for reflection on the changing nature of labor, education, self-actualization, and the concept of digital civil society (DCS), including its governance and privacy concerns |
| | | | in addition, DCS-related questions delve into the experiences and insights of civil society activists about the intersection of activism and the digitalization—covering a range of topics, from personal definitions of civil society and democracy to the practical applications and implications of digital tools on activism |

---

[3] A mixed-methods research design was employed, consisting of quantitative data collection and analysis in the first phase, followed by qualitative data collection and analysis, which were intended to explain the quantitative results in greater depth and enhance the initial findings (Creswell & Plano Clark, 2018).

For this project, I outlined and utilized what I call the method of polar opposition, which is designed to avoid various biases, including urban bias (the tendency to interview people primarily from metropolitan hubs that are easily accessible by public transportation) and liberal bias (the tendency to interview people mainly from progressive circles). Another bias that was pointed out by one of my interlocutors was what I call *negativity* bias (Participant #7, personal communication, March 24, 2024). This topic may warrant further investigation in a separate study, as scholars and scholarship alike (as well as legacy media) often focus on the negative aspects of digitalization, including numerous discussions about the detrimental impact of social media (e.g., Cocking & Van den Hoven, 2018; Corstjens & Umblijs, 2012; Elsayed, 2021), as well as the existential threats posed by AI (Hendrycks et al., 2023; Karnofsky, 2022). At the same time, some research projects strive to consider both the positive and negative aspects of technological advancement (e.g., Akram & Kumar, 2017; Kennedy, 2019). After all, the world is rarely monochromatic. Instead, it operates across the spectrum of colors and hues. Therefore, in my survey and interviews, I also asked about the positive aspects of digitalization and what makes people excited about experiencing and participating in the Fourth Industrial Revolution (4IR).

### Quantitative Phase

In the quantitative phase of this study, I collected survey data from a representative sample of participants living in the US, a technologically advanced country (*Global Innovation Index 2024 - GII 2024 Results*, n.d.; *Most Technologically Advanced Countries 2024*, n.d.), to measure their DCS participation levels as well as their attitudes and opinions toward digitalization. Questions were delivered through the *Prolific* crowdsourcing platform. *Prolific* offers a pre-screening tool that allows researchers to target specific demographics or participants from particular walks of life (Prolific, n.d.). No specific targeting was performed based on qualities other than those reflecting the US Census Bureau data (n.d.), which helped assemble the representative group of participants. Furthermore, I included a due diligence check within all surveys to help screen participants who paid little attention to detail or who might not be able to provide (reliable) answers at that moment. In all instances, final data were anonymized and securely stored in accordance with a research protocol developed and approved by the Institutional Review Board (IRB).

### Qualitative Phase

The qualitative phase was conducted to map and explain the impact of digitization on society. After obtaining IRB approval, semi-structured and open-ended interviews were conducted with a sample of 16 stakeholders. These individuals play an active role at the intersection of digitalization and civil society. The snowball interview technique used in this research enabled participants to help recruit subsequent interlocutors (Forman et al., 2008). The interviews lasted approximately 60 minutes and were fully transcribed, while *in-vivo* coding was performed during the interview (Saldaña, 2021). Depending on interlocutors' availability and preferences regarding the meeting format (either virtual or in-person), the conversation was held via *Zoom*. In an effort to overcome cognitive biases inherent in assumptions we may hold, even when posing questions (Meszaros, 2007), I also asked interlocutors if I was omitting something, and/or if there were additional questions that should have been asked. In two instances, respondents expanded on their previous answers after suggesting their own questions.

Additionally, I followed the steps for data analysis outlined by Johnny Saldaña in *The Coding Manual for Qualitative Researchers* (2021). The transcripts of interviews, after the initial *in-vivo* coding, were further coded using open and axial coding techniques (Charmaz, 2006; Saldaña, 2021). The codes were placed in thematic categories, then mapped and analyzed (Clarke & Charmaz, 2014; Saldaña, 2021).

## Findings & Discussion

Below, I discuss the descriptive statistics stemming from a survey delivered through *Prolific* to a representative sample of 348 participants (Table 1) in the US (of 377 participants, 348 provided valid responses and were included in the analysis). Data were collected on February 25, 2024, and analyzed using *R* and *Tableau*. Almost half of the respondents, 47.1% (164 participants), reported belonging to online groups (Table 2), which may serve as a proxy indicator of participation in DCS. Online group membership reflects engagement in digital spaces where individuals come together, exchange information, actively participate in discussions, or form communities of interest. Nevertheless, mere membership does not necessarily equate to active participation, though it can signal a threshold level of involvement. It is important to contextualize this finding within the sample composition, as participants were recruited from individuals who actively engage in the digital sphere, for example, through participation in *Prolific* surveys. Consequently, this cohort is unlikely to be significantly affected by the digital divide stemming from technological barriers. In comparison, AmeriCorps, a federal agency for national service and volunteerism, (*AmeriCorps*, n.d.), reported that in 2023, 25.7% of Americans reported belonging to organizations, while 49.8% reported contributed to charitable giving (*AmeriCorps*, n.d.). If DCS is to be critically examined in relation to its offline, *traditional* counterpart, the methodological foundations of such a comparison must be carefully considered. Can simply being listed as a member of an online group be compared to belonging to a parish choir if one never attends rehearsals or joins the choir's trips to Northern France—just as passive membership in an online community does not necessarily mean taking part in digital activities such as creating, signing, liking, sharing, or commenting on petitions? On the other hand, is it even methodologically sound to draw a distinction between online and offline civil society for research purposes? Whether in an authoritarian country, where joining an anti-government rally can be risky, or in a robust democracy, where protesters blocking a road to oppose laws, they consider harmful might still face police actions like pepper-spraying, acts of dissent often carry real-world consequences. *Low-effort* digital acts, when compared to those in the analog world, may also carry significant consequences. This is particularly true in non-democratic regimes. For example, in the Islamic Republic of Iran, where various individuals sentenced to capital punishment had previously posted blog entries or expressed dissent on social media against the totalitarian government (e.g., *Iran*, 2024).

Data collected for this project show that 42.8% of participants (149) actively engage in online civil society activities (Table 4). The five most popular causes that respondents reported as mobilizing them to participate in civil society were (Table 5): 1). environmental conservation and the struggle to stop the climate catastrophe (41.7%); 2). addressing unemployment, poverty, and workplace concerns (29.9%); 3). advocacy for pro-choice policies and women's rights (23.9%); 4). health and social welfare (23.3%); 5). the advancement of human rights (20.1%).

The distance between the first and second cause is quite stark, amounting to 11.8%.

**Table 4**

In the past 12-months have you engaged with online petitions: created, signed, liked, re-shared and/or commented on them? (Data from a US-based representative sample conducted on February 25th, 2024).

| response | count | % |
|---|---|---|
| yes | 149 | 42.8 |
| no | 199 | 47.1 |

**Table 5**

In your opinion, which are the issues in your country that can benefit from active civil society engagement? (Data from a US-based representative sample conducted on February 25th, 2024. Percentages may not add up to 100% as respondents could select multiple causes).

| response | count | % |
|---|---|---|
| environmental conservation and struggle to stop climate catastrophe | 145 | 41.7 |
| addressing unemployment, poverty, and workplace concerns | 104 | 29.9 |
| advocacy for pro-woman's choice policies | 83 | 23.9 |
| health and societal welfare | 81 | 23.3 |
| human rights advancement | 70 | 20.1 |
| strengthening education | 63 | 18.1 |
| racial equity | 56 | 16.1 |

To comprehensively evaluate engagement and netizenship within this research project, three indices: TCS, DCS, and PDCS were developed. These indices help guide the study by providing insight into three distinct dimensions of civil society:

- **TCS** is assessed through a composite score based on the following criteria:
  1). The range of activities in which participants have taken part, such as protests, demonstrations, strikes, membership or participation in non-profit organizations, attendance at public forums or consultations, volunteer work, involvement in grassroots initiatives, donations, event hosting, and other forms of participation outlined in the survey (0-11).
  2). A score reflecting individual and/or group associations with organizations such as labor unions, advocacy groups, charitable organizations, recreational sports clubs, cultural groups, and other voluntary associations (0-18).
  3). Whether the individual is affiliated with an institutionalized non-profit organization (0-1).

- **DCS** is assessed through a composite score based on the following criteria:

1). Participation in online groups, including those hosted on social media (0-1).
2). Involvement with online petitions, measured by whether a person has created, signed, liked, re-shared, or commented on one (0–1).
3). An engagement score for online petitions (0–5), capturing the depth of involvement in DCS—for example, whether the participant created, signed, re-shared, liked, or commented on a petition.
4). An engagement score for online platforms (0–4), indicating the number of platforms through which participants engage in digital activities.
5). An engagement score assessing whether online engagement has led to offline participation in activities such as volunteering, learning about a specific issue, attending meetings, rallies, or protests, joining a non-governmental or grassroots organization, participating in corporate social responsibility initiatives, or donating to a cause or non-profit (0–2).

- **PDCS**[4] is assessed through a composite score based on four variables:
  1). A binary indicator assessing whether the individual has engaged with political content online (0–1).
  2). An engagement score (0–6) measuring the depth of involvement, with lower scores indicating passive actions such as viewing or liking political content and higher scores reflecting active participation such as commenting, sharing, or creating political content.
  3). A binary indicator evaluating whether exposure to online information, stories, or activism has translated into offline political activism, such as attending rallies or protests (0–1).

When survey participants were asked to identify what civil society means to them (Table 6), 45.1% of the total sample defined civil society as a public sphere, and 42% of respondents indicated that civil society comprises citizen groups and other voluntary civic participants, followed by the notion of digital space (34.5%). Although perceiving civil society as a form of social and cultural capital can be a very *scholarly* way of describing it (e.g., Bourdieu, 1977; Putnam 2000), many responders also viewed civil society as such—34.5%.[5]

It is striking that when the general public encounters the notion of civil society, they understand it primarily as a concept of people and space. The problem of space is essential to understanding the current limitations of DCS. As social beings, we need Hyde Parks and Tahrir Squares and Zuccotti Parks to come together when needed, to express ourselves, to show dissent, and to engage in civil disobedience.[6] However, as much as the digital sphere can be, and often is, a platform for discursive exchange (also with various limitations), the issue comes with the realm of civil disobedience. The sphere of action can take the form of donations, likes, and/or signed petitions, but netizens rarely have the capacity to come together and peacefully disrupt. This occurs through hijacking hashtags (Bernholz, 2013), or collectively interfering with a site's ability to operate, e.g., through the Distributed Denial of Service (DDoS) attacks like those by the *Anonymous Group*. Such action would be an instance of *hacktivism*, the *mélange* of political activism and computer hacking (Denning, 1999), "the nonviolent use of illegal or legally ambiguous digital tools in pursuit of political ends" (Samuel, 2004, p.2). These tools include website defacements, redirects, denial-of-service attacks, information theft, website parodies, virtual sit-ins, virtual sabotage, and software development (Samuel, 2004). However, the DDoS actions can also be used by other agents to, for instance, impede the website of benign actors.

---

[4] It is important to add that the variety of DCS forms of organizing should not be overlooked, as politics represents only one dimension, a point underscored by the data collected and by the salience of different issues, such as animal welfare and climate catastrophe prevention, which mobilize people to engage in DCS. However, due to the nature of this research project, particular attention was given to political participation.

[5] Putnam explains, "Social capital refers here to features of social organization, such as trust, norms, and networks, that can improve the efficiency of society by facilitating coordinated actions" (Putnam et al., 1993, p. 167).

[6] Hyde Park in London embodies a long-standing tradition of free speech and public assembly, particularly at its famed Speaker's Corner, which is open to anyone wishing to express their thoughts and convictions. Tahrir Square in Cairo stands as a symbol of the Egyptian Revolution and the broader Arab Spring uprisings, while Zuccotti Park in New York City has become synonymous with the Occupy Wall Street movement, symbolizing grassroots protest against economic inequality and corporate influence in politics. However, it is also important to remember that civil society convenes not only to protest but also to celebrate, demonstrate support, and express solidarity.

**Table 6**

How would you define or describe civil society from the options given below? (Data from a US-based representative sample conducted on February 25th, 2024. Percentages may not add up to 100% as respondents could select multiple platforms).

| answer | count | % |
|---|---|---|
| a public sphere | 157 | 45.1 |
| a space made up of citizen groups and other voluntary civic participators | 146 | 42.0 |
| a digital space | 120 | 34.5 |
| a social and cultural capital of individual persons acting together | 120 | 34.5 |
| participatory democracy | 110 | 31.6 |
| a digital political and/or community-based activity | 105 | 30.2 |
| a non-governmental organization | 86 | 24.7 |
| the "third sector" primarily composed of non-governmental entities | 73 | 21.0 |
| a grassroot movement | 67 | 19.3 |
| a charity organization | 43 | 12.4 |
| a space influenced by individual donors | 38 | 10.9 |
| a non-partisan space | 35 | 10.1 |
| a space influenced by the national/federal government | 33 | 9.5 |
| a partisan space | 30 | 8.6 |
| a space influenced by the local government | 30 | 8.6 |
| corporate social responsibility (CSR) | 28 | 8.0 |
| a space influenced by international donors | 20 | 5.7 |
| a space influenced by corporate donors | 18 | 5.2 |
| unsure | 36 | 10.3 |
| other | 3 | 0.9 |

DCS, in a broader sense, also encompasses tangible support that followers of nonprofits can provide to NGOs through digital means, such as online monetary donations. However, DCS extends beyond mere financial contributions to NGOs, offering a wider range of engagement and participation opportunities. Based on participants' responses (Table 6), nonprofits were associated with civil society by 24.7% of respondents when asked to describe what the term civil society means to them. Interestingly, this perception was surpassed by that of participatory democracy, cited by 31.6% of the total sample, and digital activities, mentioned by 30.2%. Some individuals have grown disillusioned with nonprofits' ability to meaningfully address key issues (e.g., Gregory & Howard, 2009). This conviction has been reflected in different ways. 1). Many people have opted to join grassroots and digital movements, which, although often more ephemeral in nature, directly respond to urgent needs. These movements tackle specific issues without relying on private or institutional donors, paid employees, office spaces, or traditional hierarchical structures, allowing for more immediate and flexible action. 2). Many nonprofits are criticized for burning resources on operating costs or engaging in misleading activities (e.g., Arbogust, 2020; BouChabke & Haddad, 2021). 3). Nonprofits are often vulnerable to donors' needs and expectations, which can also be shaped by digitalization (Participant #14, personal communication, May 2, 2024). As my interlocutor pointed out, the digitalization of civil society can also be a double-edged sword. Consider a project scenario where a donor has invested significant resources and your team has dedicated months of labor, only for the outcome to be a handful of likes on social media. In the past, producing a well-crafted video that conveyed a compelling story, supported by data and testimonies of injustice, might have been considered a meaningful achievement. However, in today's digital landscape, if that same video receives only three likes and one re-share, it can hardly be regarded as a success, even though its digital presence allows the impact to be at least quantified.

Moreover, as my Interlocutor, a nonprofit leader, emphasized, the evolving landscape of digital regulations can significantly impact how NGOs operate, for better or worse. For instance, new regulations may impose additional burdens and costs on nonprofits by complicating how they store and process data (Participant #14, personal communication, May 2, 2024). However, this is a cost that institutionalized civil society must consider, just as it does with other aspects of its operations, to ensure that its beneficiaries are protected and not put at risk by nonprofit activities themselves, for example, through the inadvertent disclosure of private or sensitive data. On the other hand, the digital footprint of visitors to nonprofit websites can also be seen as a form of data donation, which exists alongside more advanced forms of data contribution, such as donating biomedical or fitness data for medical research purposes (Abdelhamid, 2021).

The salient issue arises with participants' perception of the effectiveness of DCS (Table 7). For the majority (59.2%), it is seen as mere *slacktivism,* often reduced to online petitions (53.2%). Meanwhile, 33.9% view it as a movement, while 31.3% perceive it as a form of effective action. DCS has carried the label of *slacktivism* attached to it since its early days (e.g., Reardon, 2013). Today, especially in the Global North, being offline has become something of a privilege, and DCS has simply emerged as a default way of engagement with the world, a primary way people interact with one another and construct both civil society and their own sense of netizenship.

Another question remains: *Does the DCS necessitate any analog follow-up?* Based on the survey answers (Table 8), in the past 12 months, respondents who felt mobilized to act based on online content (125 participants, 35.9% of the overall sample) expressed that thanks to DCS (Table 9) they had learned more about a particular issue (83.2%) or donated money (49.6%). These two acts were followed by joining an informal group (17.6%), volunteering (16.8%), or physically attending a rally/protest (14.4%) or a meeting (11.2%). Interestingly, 9.6% joined a nonprofit because of DCS, highlighting both the untapped potential of NGOs to channel digital energy into their membership beyond their informational and fundraising efforts. The survey results revealed that 12% of participants (42 out of 348) actively engaged in an offline follow-up to DCS (Table 10). Among these 42 participants, 44% indicated participation in an event/meeting, and 36% indicated involvement in a community service event or rally/protest (Table 11). Only 2.3% of the overall sample had ever created a petition that could be signed, liked, or re-shared by others (Table 12). In fact, I have long grappled with the question of whether DCS requires a corresponding manifestation in the offline, analog sphere and whether the absence of such offline follow-up should justify a conceptual distinction between more active forms of DCS and phenomena commonly labeled as *slacktivism*. However, I am hesitant to impose a rigid dichotomy, as even within *traditional*, analog civil society, passive forms of participation exist, such as paying membership dues without active engagement. Yet the mere fact of holding a membership card (e.g., of the American Civil Liberties Union) contributes to a sense of belonging to a collective larger than oneself. On the flip side, how many individuals actively organize events or protests in the offline sphere? The distribution of truly engaged participation in digital and *traditional* civil society may resemble each other, which warrants further exploration and research. Moreover, in the future, this study could greatly benefit from a longitudinal analysis to assess, for example, the impact of Donald J. Trump's second presidency on DCS in the US.

**Table 7**

What comes to your mind when you hear the word digital civil society (DCS)—occurrences and mobilizations of civil society taking place online, e.g., on social media, web-forums, digital communication apps, or initiating online and continuing in the offline public sphere (Data from a US-based representative sample conducted on February 25th, 2024. Percentages may not add up to 100% as respondents could select multiple platforms).

| answer | count | % |
|---|---|---|
| slacktivism (minimal effort) | 206 | 59.2 |
| online petitions | 185 | 53.2 |
| movement | 118 | 33.9 |
| effective action | 109 | 31.3 |
| other | 40 | 11.5 |

**Table 8**

In the past 12-months, have you felt mobilized to act based on online content?
(Data from a US-based representative sample conducted on February 25th, 2024).

| | count | % |
|---|---|---|
| yes | 125 | 35.9 |
| no | 223 | 64.1 |

**Table 9**

In the past 12-months, have you felt mobilized to act based on online content? Please specify. (Data from a US-based representative sample conducted on February 25th, 2024. Percentages may not add up to 100% as respondents could select multiple platforms).

| answer | count | % |
|---|---|---|
| I learned more about particular issue | 104 | 83.2 |
| donated money | 62 | 49.6 |
| joined an informal group | 22 | 17.6 |
| volunteered | 21 | 16.8 |
| attended a rally/protest | 18 | 14.4 |
| attended a meeting | 14 | 11.2 |
| joined a nonprofit/NGO | 12 | 9.6 |

**Table 10**

In the past 12-months have you joined any offline event, meeting, town hall meeting, rally/protest, for which you were invited/informed about online? (Data from a US-based representative sample conducted on February 25th, 2024).

| | count | % |
|---|---|---|
| yes | 42 | 12.1 |
| no | 306 | 87.9 |

**Table 11**

In the past 12-months have you joined any offline event, meeting, town hall meeting, rally/protest, for which you were invited/informed about online? Please specify. Data from a US-based representative sample conducted on February 25th, 2024. *Below percentages represent the subset of respondents who attended offline events they learned about online, which was 12.1% of the total sample.*

| answer | count | % of subset | % of total sample |
|---|---|---|---|
| meeting/event | 22 | 44 | 6.3 |
| community service rally/protest | 18 | 36 | 5.2 |
| town hall meeting | 10 | 20 | 2.9 |

**Table 12**

In the past 12 months, have you engaged with online petitions: created, signed, liked, re-shared and/or commented on them? (N=348) (Data from a US-based representative sample conducted on February 25th, 2024. Percentages may not add up to 100% as respondents could select multiple platforms).

| answer | count | % of total sample |
|---|---|---|
| signed | 122 | 35.1 |
| liked | 80 | 23 |
| re-shared | 55 | 15.8 |
| commented | 44 | 12.6 |
| created one | 8 | 2.3 |
| other | 1 | 0.3 |

Furthermore, when the survey data were dissected by age (Table 13), the highest average score for participation in the TCS was found among participants aged 60 to 65 (7.9) and 72 to 77 (7.8)[7]; for DCS, the highest scores were among 19–23-year-olds (3.4) and 54–59-year-olds (3.4). The PDCS yielded the highest average among the youngest cohort surveyed, those aged 19-23 (2.3); followed by 24–29-year-olds (2.3), and 54–65-year-olds (2.2). These findings are consistent with the data reported by *AmeriCorps Civic Engagement* measure of organizational membership, which shows that the age group of 70-79 years old had the highest rate (34.2%) of membership in TCS, whereas individuals aged 18 and 24 recorded the highest rate of participation (21.4%) in sharing their views online (*AmeriCorps Civic Engagement and Volunteering Dashboard*, n.d.).

Among socio-economic and political ideologies (Table 14), those who self-identified as "very progressive" scored the highest across the three measures of TCS, DCS, and PDCS (8.5/4.3/2.9, respectively), followed by those who self-identified as "progressive" (8.3/4.0/2.8, respectively). Apart from DCS, where individuals who self-identified as "other" recorded the third-highest average (3.5), the third-highest scores for TCS and PDCS were reported among those who self-identified as "very conservative" (7.5 and 2.5, respectively). These results speak to the polarized nature of today's political landscape, where those on opposing sides of the political spectrum are also involved and active but remain vulnerable to placement in social media echo chambers, where algorithm-driven content can reinforce existing societal divisions. In addition, it highlights the untapped potential for the liberal left to make greater use of the digital sphere for their political rallying, something that has been well used by the conservatives, e.g., Truth Social. Among party affiliations (Table 15), those who scored the highest averages were Democrats (7.5 for TCS, 3.6 for DCS, and 2.4 for PDCS), followed by those who identified as other for the TCS (7.1) and DCS (3.4), and by Independents for PDSC. Moreover, 47.1% of surveyed respondents (164 participants) indicated that over the past 12 months, they had engaged with online political content, which underscores and reinforces the argument that the digital sphere functions as an inherently political space (Table 16). Such engagement included creating, liking, re-sharing, or commenting on political content online. However, similar to the general participation of netizens in DCS (Table 17), the active and involved act of creating political content online was the fourth most popular activity reported by participants (4% of those who engaged in PDCS), lagging behind simply liking political content (39.4%), commenting (29.1%), and re-sharing (24.4%).

Finally, among 151 participants who had engaged with online petitions in the past 12 months—whether by creating, signing, liking, re-sharing, or commenting on them—DCS engagement was primarily experienced through Meta platforms, which ranked first and third most popular options. Specifically, 63.6% of participants used Facebook and 34.4% used Instagram (Table 18). The second most popular platform was X (formerly Twitter), used by 35.1% of participants, while the fourth was TikTok (14.2%).

---

[7] The age groups were construed in five-year intervals.

**Table 13**

Summary table comparing TCS, DCS, and PDCS scores across different age groups.

|  | average of TCS_score | average of DCS_score | average of PDCS_score |
|---|---|---|---|
| *19-23* | 7.3 | 3.4 | 2.3 |
| *24-29* | 7.0 | 3.3 | 2.3 |
| *30-35* | 5.5 | 2.8 | 1.7 |
| *36-41* | 6.8 | 3.1 | 1.8 |
| *42-47* | 7.3 | 3.1 | 1.8 |
| *48-53* | 7.0 | 3.1 | 2.2 |
| *54-59* | 6.8 | 3.4 | 2.2 |
| *60-65* | 7.9 | 3.2 | 2.2 |
| *66-71* | 6.8 | 2.6 | 1.6 |
| *72-77* | 7.8 | 2.9 | 2.0 |
| *78-83* | 3.5 | 1.0 | 1.0 |
| *average* | *7.0* | *3.1* | *2.0* |

**Table 14**

Summary table comparing TCS, DCS, and PDCS scores across different political ideologies.

|  | average of TCS_score | average of DCS_score | average of PDCS_score |
|---|---|---|---|
| very conservative | 7.5 | 1.9 | 2.5 |
| conservative | 6.5 | 2.6 | 1.8 |
| somewhat conservative | 6.2 | 2.6 | 1.5 |
| someone in between | 5.8 | 2.3 | 1.2 |
| somewhat progressive | 6.9 | 2.8 | 1.5 |
| progressive | 8.3 | 4.0 | 2.8 |
| very progressive | 8.5 | 4.3 | 2.9 |
| neither | 2.9 | 1.8 | 1.1 |
| other | 1.5 | 3.5 | 2.0 |
| *average* | *7.0* | *3.1* | *2.0* |

**Table 15**

Summary table comparing TCS, DCS, and PDCS scores across different political affiliations.

|  | average of TCS_score | average of DCS_score | average of PDCS_score |
|---|---|---|---|
| Republican | 6.3 | 2.1 | 1.6 |
| Democrat | 7.5 | 3.6 | 2.4 |
| Independent | 7.0 | 3.2 | 2.0 |
| none | 3.0 | 1.5 | 0.7 |
| other | 7.1 | 3.4 | 1.3 |
| *average* | *7.0* | *3.1* | *2.0* |

**Table 16**

In the past 12-months have you created, liked, re-shared and/or commented on political content online?

|  | count | % |
|---|---|---|
| yes | 164 | 47.1 |
| no | 184 | 52.9 |

**Table 17**

In the past 12-months have you created, liked, re-shared and/or commented on political content online? Pinned to homepage.  Please specify. (Data from a US-based representative sample conducted on February 25[th], 2024. Percentages may not add up to 100% as respondents could select multiple platforms).

| answer | count (n=164) | % of subset | % of total sample |
|---|---|---|---|
| liked political content online | 137 | 83.5 | 39.4 |
| commented on political content online | 101 | 61.6 | 29.0 |
| re-shared political content online | 85 | 51.8 | 24.4 |
| created political content online | 14 | 8.5 | 4.0 |
| other | 5 | 3.0 | 1.4 |

**Table 18**

Please indicate the platform of your digital engagement—a question asked to the 151 participants who in the past 12-months engaged with online petitions (Data from a US-based representative sample conducted on February 25th, 2024. Percentages may not add up to 100% as respondents could select multiple platforms).

| answer | count | % |
|---|---|---|
| Facebook | 96 | 63.6 |
| X (formerly: Twitter) | 53 | 35.1 |
| Instagram | 52 | 34.4 |
| TikTok | 40 | 26.5 |
| other | 38 | 25.2 |

## Conclusion

To conclude, DCS represents a *new normal* for civil society, serving not merely as an extension or a step in its evolution, but as its contemporary embodiment. There are already generations of Americans who know civil society mostly through the digital realm, and as indicated by recent *Pew Research*, 41% of Americans report being constantly online (Fetterolf, 2025). Nonetheless, the most pressing issue, I argue, rests in the way the digital public sphere is owned and regulated, in order to allow people to *bowl online* as freely and as much as they see is necessary. The absence of a clear conceptualization or operationalization of the DPS presents a substantial, potentially existential, challenge to DCS in the long run, especially as society faces increasing grievances against big tech companies (Lichtinger & Hosseini Maasoum, 2025). As our civic engagement and social organizing take place increasingly, and in many cases, predominantly, online, the next step is for us to conceptually articulate DPS, our cyber Tahrir Squares and Zuccotti Parks, and to technologically materialize them. The comprehensive conceptualization and operationalization of DCS could provide democratic societies with a structured pathway to address the multifaceted challenges posed by the digital era. While social science still wrestles with how to conceptualize and fully grasp DCS, civil society has already become digital, with netizens affirming that premise daily, *clicking* to create the change they wish to see in the world. They are *bowling online*!

## Acknowledgments

I would like to express my gratitude To My Wife Liqian, My Parents, Teresa and Walenty, and to every Upstander of the Digital Era who has inspired this work. I am also grateful to Prof. Ari Kohen, Prof. Adrian Wisnicki, and Prof. Ingrid Haas for helping me excel in academia and for introducing me to the world of political philosophy (Prof. Kohen), digital humanities (Prof. Wisnicki), and technological governance and public opinion/attitudes (Prof. Haas, with whom I successfully obtained a grant that funded this project). Thanks to them, along with Prof. Ross Miller and Prof. Wayne Babchuk, I have been able to stand on the shoulders of giants. Moreover, I am thankful to the Nebraska Governance and Technology Center (Prof. Gus Hurwitz, Ms. Cindy Harris), which provided generous support and funding for the data collection that made this this research possible. In addition, I would like to thank Prof. Kristin Goss, the chair of the panel "Power to the People? Civil Society, Power, and Powerlessness," at the American Political Science Association Annual Meeting in 2025, and Prof. Natalia Bueno, the discussant of my paper, for their valuable feedback. Finally, I am grateful to Prof. Lucy Bernholz and Prof. Toussaint Nothias for their seminal contributions to the study of digital civil society.

## Competing Interests

The author has no relevant financial or non-financial interests to disclose.

## Funding

This work was generously supported by the Nebraska Governance and Technology Center (Grant Number: 27-1502-0189-015), Co-Principial Investigator: Prof. Ingrid J. Haas.

## Ethical Approval

IRB Project ID#: 20240222618EP. Study Title: Digital Society: opinions, attitudes, and participation.

## Works Cited

Abdelhamid, M. (2021). Fitness tracker information and privacy management: Empirical study. *Journal of Medical Internet Research*, *23*(11).

*Akcja Demokracja*. (n.d.). Retrieved July 18, 2024, from https://www.akcjademokracja.pl/

Akram, W., & Kumar, R. (2017). A study on positive and negative effects of social media on society. *International Journal of Computer Sciences and Engineering*, *5*(10), 351–354.

Alscher, M., Priller, E., Ratka, S., & Strachwitz, R. G. (2017). *The Space for Civil Society: Shrinking? Growing? Changing?*

Alvarez, S. E. (2009). Beyond NGO-ization?: Reflections from Latin America. *Development*, *52*(2), 175–184.

*AmeriCorps Civic Engagement and Volunteering (CEV) Dashboard*. (n.d.). Retrieved February 3, 2025, from https://data.americorps.gov/stories/s/AmeriCorps-Civic-Engagement-and-Volunteering-CEV-D/62w6-z7xa/

Aragón, P., Sáez-Trumper, D., Redi, M., Hale, S., Gómez, V., & Kaltenbrunner, A. (2018). *Online petitioning through data exploration and what we found there: A dataset of petitions from avaaz. Org*. twelfth international AAAI conference on web and social media.

Arbogust, M. (2020). *Why Do Nonprofits Fail? A Quantitative Study of Form 990 Information in the Years Preceding Closure*.

Bal, A. S., Archer-Brown, C., Robson, K., & Hall, D. E. (2013). Do good, goes bad, gets ugly: Kony 2012. *Journal of Public Affairs*, *13*(2), 202–208.

Barber, B. (2003). *Strong democracy: Participatory politics for a new age*. Univ of California Press.

Berman, S. (1997). Civil society and the collapse of the Weimar Republic. *World Politics*, *49*(3), 401–429.

Bermeo, N. (2016). On Democratic Backsliding. *Journal of Democracy*, *27*(1), 5–19. https://doi.org/10.1353/jod.2016.0012

Bernholz, L. (2013) Inventing Digital Civil Society - YouTube. (n.d.). Retrieved October 22, 2023, from https://www.youtube.com/

Bernholz, L., Cordelli, C., & Reich, R. (2013). *The emergence of digital civil society*. Stanford Center on Philanthropy and Civil Society.

Boswell, E., & Babchuk, W. A. (2023). *Philosophical and theoretical underpinnings of qualitative research*. Elsevier.

BouChabke, S., & Haddad, G. (2021). Ineffectiveness, Poor Coordination, and Corruption in Humanitarian Aid: The Syrian Refugee Crisis in Lebanon. *VOLUNTAS: International Journal of Voluntary and Nonprofit Organizations*, *32*(4), 894–909. https://doi.org/10.1007/s11266-021-00366-2

Bourdieu, P. (1977). *Reproduction in Education: Society and Culture*. Sage Publications.

Bourdieu, P. (1985). The forms of capital. In *Handbook of Theory and Reserch for the Sociology of Eduction ed. John G. Richrdson* (Vol. 1, pp. 241–258). Greenwood.

Charmaz, K. (2006). *Constructing grounded theory: A practical guide through qualitative analysis*. sage.

Chenoweth, E., & Stephan, M. J. (2011). *Why Civil Resistance Works: The Strategic Logic of Nonviolent Conflict*. Columbia University Press.

Citizen Lab. (2018, September 18). *Hide and Seek: Tracking NSO Group's Pegasus Spyware to Operations in 45 Countries - The Citizen Lab*. https://citizenlab.ca/2018/09/hide-and-seek-tracking-nso-groups-pegasus-spyware-to-operations-in-45-countries/

Clarke, A. E., & Charmaz, K. (2014). *Grounded theory and situational analysis*. Sage.

Cocking, D., & Van den Hoven, J. (2018). *Evil online* (Vol. 15). John Wiley & Sons.

Corstjens, M., & Umblijs, A. (2012). The power of evil: The damage of negative social media strongly outweigh positive contributions. *Journal of Advertising Research*, *52*(4), 433–449.

Cover, R., Thompson, J. D., & Haw, A. (2022). The spectre of populist leadership: QAnon, emergent formations, and digital community. *Media and Communication*, *10*(4), 118–128.

Creswell, J. W., & Plano Clark, V. L. (2018). *Designing and conducting mixed methods research*.

Denning, D. (1999). *Activism, hacktivism, and cyberterrorism: The internet as a tool for influencing foreign policy [On-line]*. Networks and Netwars: The Future of Terror, Crime, and Military, Rand, Santa Monica, disponibil la http://www. nautilus. org/infopolicy/workshop/papers/denning. html.

Diamond, L. (1994). Rethinking civil society: Toward democratic consolidation. *Journal of Democracy*, *5*(3), 4–17.

Diamond, L. (2019). The road to digital unfreedom: The threat of postmodern totalitarianism. *Journal of Democracy*, *30*(1), 20–24.

Elsayed, W. (2021). The negative effects of social media on the social identity of adolescents from the perspective of social work. *Heliyon*, *7*(2).

Emery, F. E., & Trist, E. L. (1960). Socio-technical systems. *Management Science, Models and Techniques*, *2*, 83–97.

Fetterolf, J. (2025, September 8). Most adults across 24 countries are online at least several times a day. *Pew Research Center*. https://www.pewresearch.org/short-reads/2025/09/08/most-adults-across-24-countries-are-online-at-least-several-times-a-day/

Floridi, L. (2009). Philosophical conceptions of information. In *Formal theories of information: From Shannon to semantic information theory and general concepts of information* (pp. 13–53). Springer.

Forman, J., Creswell, J. W., Damschroder, L., Kowalski, C. P., & Krein, S. L. (2008). Qualitative research methods: Key features and insights gained from use in infection prevention research. *American Journal of Infection Control*, *36*(10), 764–771.

Gleason, B. (2013). # Occupy Wall Street: Exploring informal learning about a social movement on Twitter. *American Behavioral Scientist*, *57*(7), 966–982.

*Global Innovation Index 2024—GII 2024 results*. (n.d.). Retrieved February 19, 2025, from https://www.wipo.int/web-publications/global-innovation-index-2024/en/gii-2024-results.html

Gregory, A. G., & Howard, D. (2009). The nonprofit starvation cycle. *Stanford Social Innovation Review*, *7*(4), 49–53.

Habermas, J. (1991). *The Structural Transformation of the Public Sphere: An Inquiry into a Category of Bourgeois Society*. MIT Press.

Habermas, J. (2006). Political communication in media society: Does democracy still enjoy an epistemic dimension? The impact of normative theory on empirical research. *Communication Theory*, *16*(4), 411–426.

Hendrycks, D., Mazeika, M., & Woodside, T. (2023). An overview of catastrophic ai risks. *arXiv Preprint arXiv:2306.12001*.

*Home | AmeriCorps*. (n.d.). Retrieved February 19, 2025, from https://americorps.gov/

*Iran: UN experts alarmed by death sentence imposed on peaceful activist, demand moratorium on death penalty*. (2024, May 13). OHCHR. https://www.ohchr.org/en/press-releases/2024/05/iran-un-experts-alarmed-death-sentence-imposed-peaceful-activist-demand

Ivankova, N. V., Creswell, J. W., & Stick, S. L. (2006). Using mixed-methods sequential explanatory design: From theory to practice. *Field Methods*, *18*(1), 3–20.

Jadczak, S., & Słowik, P. (2025, May 15). *Ujawniamy. Ingerencja w wybory, spoty bez autora i Akcja Demokracja*. WP Wiadomości. https://wiadomosci.wp.pl/ujawniamy-ingerencja-w-wybory-spoty-bez-autora-i-akcja-demokracja-7156892271278624a

Karnofsky, H. (2022, June 9). *AI Could Defeat All Of Us Combined*. Cold Takes. https://www.cold-takes.com/ai-could-defeat-all-of-us-combined/

Kennedy, K. (2019). *Positive and negative effects of social media on adolescent well-being*. Minnesota State University, Mankato.

Kowal, D. M. (2002). Indigenous Voices: The Zapatista Movement. *Critical Perspectives on the Internet*, 105.

Lanier, J. (2014). *Who Owns the Future?* Simon & Schuster.

Lanier, J. (2010). *You are not a gadget*. Alfred A. Knopf.

Lichtinger, G., & Hosseini Maasoum, S. M. (2025). Generative AI as Seniority-Biased Technological Change: Evidence from US Resume and Job Posting Data. *Available at SSRN*.

Lombardo, G., Mordonini, M., & Tomaiuolo, M. (2021). Adoption of social media in socio-technical systems: A survey. *Information*, *12*(3), 132.

Lührmann, A., & Lindberg, S. I. (2019). A third wave of autocratization is here: What is new about it? *Democratization*, *26*(7), 1095–1113. https://doi.org/10.1080/13510347.2019.1582029

Meszaros, G. (2007). Researching the landless movement in Brazil. *Research Methods for Law*, 133–158.

Mossberger, K., Tolbert, C. J., & McNeal, R. S. (2007). *Digital citizenship: The Internet, society, and participation*. MIt Press.

*Most Technologically Advanced Countries 2024*. (n.d.). World Population Review. Retrieved April 23, 2024, from https://worldpopulationreview.com/country-rankings/most-technologically-advanced-countries

Mundt, M., Ross, K., & Burnett, C. M. (2018). Scaling social movements through social media: The case of Black Lives Matter. *Social Media+ Society*, *4*(4), 2056305118807911.

Participant #7. (2024, March 24). *Interview #7* [Personal communication].

Participant #8. (2024, March 24). *Interview #8* [Personal communication].

Participant #14. (2024, May 2). *Interview #14* [Personal communication].

Pateman, C. (1970). *Participation and democratic theory*. Cambridge University Press.

Prainsack, B. (2019). Data donation: How to resist the iLeviathan. *The Ethics of Medical Data Donation*, 9–22.

*Prolific · Quickly find research participants you can trust.* (n.d.). Retrieved October 23, 2023, from https://www.prolific.com/

Putnam, R. D. (2000). *Bowling alone: The collapse and revival of American community*. Simon & Schuster.

Putnam, R. D., Leonardi, R., & Nanetti, R. (1993). *Making democracy work: Civic traditions in modern Italy*. Princeton Univ. Press.

Reardon, S. (2013). *Does online "slacktivism" actually do more harm than good?*

Risse, M. (2021). The Fourth Generation of Human Rights: *Carr Center for Human Rights Policy Harvard Kennedy School*, 25.

Risse, M. (2023). *Political Theory of the Digital Age: Where Artificial Intelligence Might Take Us*. Cambridge University Press.

Roche, C., & Hewett, A. (2013). *The end of the golden age of international NGOs*. 21–22.

Rothschild, V. (2024). Protest and repression in China's digital surveillance state. *Journal of Information Technology & Politics*, 1–16. https://doi.org/10.1080/19331681.2024.2326464

Rotman, D., Vieweg, S., Yardi, S., Chi, E., Preece, J., Shneiderman, B., Pirolli, P., & Glaisyer, T. (2011). *From slacktivism to activism: Participatory culture in the age of social media*. 4.

Roy, A. (2014). The NGO-ization of Resistance. *Massalijn News. Http://Massalijn.Nl/New/the-Ngo-Ization-of-Resistance (Accessed 23 January 2020)*.

Saldaña, J. (2012). *The Coding Manual for Qualitative Researchers*. SAGE Publications.

Samuel, A. W. (2004). *Hacktivism and the future of political participation*. Harvard University.

Schäfer, M. S. (2015). Digital public sphere. *The International Encyclopedia of Political Communication*, *15*, 1–7.

Sony, M., & Naik, S. (2020). Industry 4.0 integration with socio-technical systems theory: A systematic review and proposed theoretical model. *Technology in Society*, *61*, 101248. https://doi.org/10.1016/j.techsoc.2020.101248

Sowa, J. (2015). Goldex Poldex Madafaka, czyli raport z (oblężonego) Pi sektora. *W: Europejskie Polityki Kulturalne*.

Tusalem, R. F. (2007). A boon or a bane? The role of civil society in third-and fourth-wave democracies. *International Political Science Review*, *28*(3), 361–386.

US Census Bureau. (n.d.). *Data*. Census.Gov. Retrieved May 4, 2024, from https://www.census.gov/data

Van Dijk, J. (2020). *The digital divide*. John Wiley & Sons.

Vie, S. (2014). In defense of "slacktivism": The Human Rights Campaign Facebook logo as digital activism. *First Monday*.

Wasko, M. M., & Faraj, S. (2005). Why should I share? Examining social capital and knowledge contribution in electronic networks of practice. *MIS Quarterly*, 35–57.

Zhang, X., Lyu, H., & Luo, J. (2021). What contributes to a crowdfunding campaign's success? Evidence and analyses from GoFundMe data. Journal of Social Computing, 2(2), 183–192.

Zuboff, Shoshana. (2017). *The Age of Surveillance Capitalism*.